\pdfoutput=1
\documentclass[11pt,a4paper]{article}
\usepackage[hyperref]{acl2021}
\usepackage{times}
\usepackage{latexsym}

\usepackage{amsmath,graphicx}

\usepackage{pgfplots}
\usepackage{lipsum,adjustbox}
\pgfplotsset{compat=1.16}
\usepackage{textcomp}

\usepackage{microtype}

\aclfinalcopy 

\newcommand{\mg}[1]{{\textcolor{black}{#1}}}

\title{Beyond Voice Activity Detection: \\
Hybrid Audio Segmentation for Direct Speech Translation}

\author{Marco Gaido\textsuperscript{$\dagger$, $*$}, Matteo Negri\textsuperscript{$\dagger$}, Mauro Cettolo\textsuperscript{$\dagger$}, Marco Turchi\textsuperscript{$\dagger$} \\
  \textsuperscript{$\dagger$}Fondazione Bruno Kessler \\
  \textsuperscript{$*$}University of Trento \\
  \texttt{\{mgaido,cettolo,negri,turchi\}@fbk.eu} \\}

\date{}

\begin{document}
\maketitle
\begin{abstract}
The audio segmentation  mismatch between training data and those seen at run-time is a major problem in direct speech translation. Indeed, while systems are usually trained on manually segmented corpora, in real use cases they are often presented with continuous audio requiring  automatic (and sub-optimal) segmentation. After comparing existing techniques (VAD-based, fixed-length  and hybrid segmentation methods), in this paper we propose enhanced hybrid solutions to produce better results without sacrificing latency. Through experiments on different domains and language pairs, we show that our methods outperform all the other techniques, reducing by at least 30\% the gap between the traditional VAD-based approach and optimal manual segmentation.
\end{abstract}

\section{Introduction}
Speech-to-text translation (ST) consists in translating  utterances in one language  into text in another language. From the architectural standpoint, ST systems are traditionally divided in cascade and direct. Cascade solutions  first transcribe the audio via automatic speech recognition (ASR) and then translate the generated transcripts with a machine translation (MT) component. In direct ST, a single end-to-end model operates without intermediate representations. This  allows reducing error propagation and latency, as well as exploiting more information (e.g. speaker’s vocal traits and prosody).

Different from MT, where sentence-level splits represent a  natural (though not necessarily optimal) input segmentation criterion,  handling audio data is more problematic. Existing  training corpora \cite{CATTONI2021101155,europarlst} split  continuous speech  into  utterances according to strong punctuation marks in the transcripts  (which are known in advance), reflecting linguistic criteria related to sentence well-formedness. This (\textit{manual}) segmentation is optimal, as it allows  ST systems to potentially generate correct outputs even for languages with different syntax and word order (e.g. subject-verb-object vs subject-object-verb). At run-time, though, audio transcripts are not known in advance and \textit{automatic} segmentation techniques have to be applied.  The traditional approach is to adopt a Voice Activity Detection (VAD) tool to break the audio on speaker silences \cite{sohn_vad},  considered as  a proxy of clause boundaries. However, since the produced segmentation is not driven by syntactic information (unlike that of the training corpora), final performance on downstream tasks  degrades considerably \cite{sinclair_segm}.

The impact of a syntax-unaware segmentation can be limited in cascade systems by means of dedicated components  that re-segment the ASR transcripts,  so to feed  MT with well-formed sentences \cite{matusov_segm}.  
%
The absence of intermediate transcripts makes this solution unfeasible for direct systems, whose performance is therefore highly sensitive to sub-optimal audio segmentation. This has been shown in the
\mg{2020} IWSLT evaluation campaign  \cite{ansari-etal-2020-findings}, where  the best direct ST system had a key feature in the  segmentation algorithm \cite{potapczyk-przybysz-2020-srpols}. 
In the same evaluation setting, the second-best direct system \cite{apptek_2020} exploited an external ASR model to segment the audio  (with a +10\% BLEU gain  compared to its VAD-based counterpart). This solution, however, formally makes it closer to a cascade architecture, losing the advantage of the reduced latency of direct systems. For this reason, while in $\S\ref{sec:results}$ we  compare with the state-of-the-art method proposed in \cite{potapczyk-przybysz-2020-srpols}, we will not consider approaches needing additional models (e.g. ASR) like the one in \cite{apptek_2020}.

So  far, no work analyzed in depth the strengths and weaknesses of different audio segmentation methods in the context of direct ST. To fill this gap,
 we study the behavior of the existing techniques and, based on the resulting observations, we propose improved hybrid methods that can also be applied to streaming audio.  Through experiments in two domains  (TED and  European Parliament  talks) and two target languages (German and Italian), we show that our solutions outperform the others in all conditions, reducing the gap with optimal manual segmentation by  at least 30\% compared to VAD systems.
 
\section{Audio Segmentation Methods}

\subsection{Existing Methods}

\noindent \textbf{VAD systems.} VAD tools are classifiers that determine whether a given audio frame contains speech or not. Based on this, a VAD-based segmentation considers a sequence of consecutive speech frames as a  segment, filtering out non-speech frames. In this work, we evaluate two widely used open source VAD tools: LIUM~\cite{meigner2010lium} and WebRTC's VAD.\footnote{http://webrtc.org/. We use the Python interface http://github.com/wiseman/py-webrtcvad.}
For LIUM, we apply the configuration employed in the IWSLT campaign \cite{ansari-etal-2020-findings}.
WebRTC takes as parameters the \textit{frame size} (10, 20 or 30ms) and the \textit{aggressiveness mode}  (an integer in the range [0, 3], 3 being the most aggressive). We select three configurations  based on the segmentation they produce on the MuST-C test set \cite{CATTONI2021101155}, one of the test sets used in our experiments (see $\S$\ref{sec:results}).   Specifically, we consider those not generating too many (more than two times the segments of the manual segmentation) or too long segments (more than 60s). They are: \textit{(3, 30ms)}, \textit{(2, 20ms)} and \textit{(3, 20ms)}. The statistics computed for the two  segmentation tools  on the MuST-C test set are presented in Table \ref{tab:vad_summary}, along with those corresponding to manual segmentation. To better understand the impact of different VADs on  translation quality, the tools are compared  on MuST-C and Europarl-ST data. Table \ref{tab:vad_res} reports preliminary translation results for  English-German (en-de) and English-Italian (en-it), obtained with the systems described in $\S\ref{sec:expsett}$. LIUM and the most aggressive WebRTC configuration \textit{(3, 20ms)} are significantly worse than the other two WebRTC configurations.
As \textit{(2, 20ms)} achieves comparable BLEU performance to \textit{(3, 30ms)} on  MuST-C and better on Europarl-ST, it is used in the rest of the paper. 

\begin{table}[htbp]
    \centering
    \small
    \centering
    \begin{tabular}{l|c|c|ccc}
    \hline
    \textbf{System} & Man. & LIUM & \multicolumn{3}{c}{WebRTC} \\
    \textbf{Aggress.} & ~~~ & ~~~ & 3 & 2 & 3  \\
    \textbf{Frame size} & ~~~ & ~~~ & 30ms & 20ms & 20ms  \\
    \hline
        \textbf{\% filtered}  & 14.66 & 0.00 & 11.27 & 9.53 & 15.58 \\
        \textbf{Num segm.}  & 2,574 & 2,725 & 3,714 & 3,506 & 5,005 \\
        \textbf{Max len (s)}  & 51.97 & 18.63 & 48.84 & 58.62 & 46.76 \\
        \textbf{Min len (s)}  & 0.05 & 2.50 & 0.60 & 0.40 & 0.40 \\ 
        \textbf{Avg len (s)}  & 5.82 & 6.44 & 4.19 & 4.53 & 2.96 \\   
    \hline
    \end{tabular}
    \caption{\label{tab:vad_summary} Statistics for  different segmentations of the MuST-C test set. 
    ``Man.'' = sentence-based segmentation.
    ``\% filtered'' =  percentage of audio discarded.}
\end{table}

 {
 \setlength{\tabcolsep}{1.5pt}

 \begin{table}[htpb]
    \centering
    \small
    \begin{tabular}{l|cccc}
    \hline
    \textbf{VAD System} &
    \multicolumn{2}{c}{\textbf{MuST-C}} & \multicolumn{2}{c}{\textbf{Europarl-ST}} \\
    & \textbf{BLEU ($\uparrow$)} & \textbf{TER ($\downarrow$)} & \textbf{BLEU ($\uparrow$)} & \textbf{TER ($\downarrow$)} \\
    \hline
    \multicolumn{5}{c}{\textbf{English-German}} \\
    \hline
    LIUM          & 19.55  & 76.21  & 15.39 & 94.06 \\
    WebRTC 3, 30ms & \textbf{21.90}  & 66.96  & 16.23 & 89.35 \\
    WebRTC 3, 20ms & 19.48  & 72.25  & 14.07 & 99.32 \\
    WebRTC 2, 20ms & 21.87  & \textbf{66.72}  & \textbf{18.51} & \textbf{78.12} \\
    \hline
    \multicolumn{5}{c}{\textbf{English-Italian}} \\
    \hline
    LIUM          & 21.29 & 67.50 & 18.88 & 73.73 \\
    WebRTC 3, 30ms & \textbf{22.46} & \textbf{64.99} & 19.85 & 72.28 \\
    WebRTC 3, 20ms & 20.09 & 68.62 & 17.35 & 78.18 \\
    WebRTC 2, 20ms & 22.34 & 66.12 & \textbf{20.90} & \textbf{69.54} \\
    \hline
    \end{tabular}
    \caption{\label{tab:vad_res} 
    Results of the VAD systems on MuST-C and Europarl-ST for en-de and en-it.}
\end{table}
}

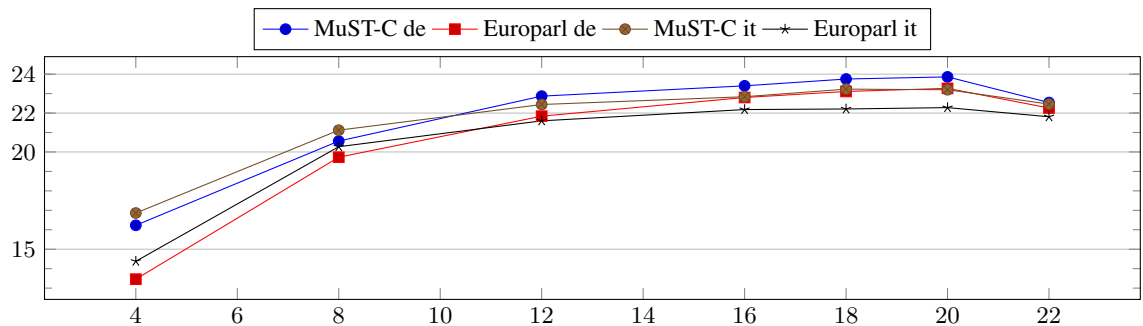
\begin{figure*}[!ht]
\centering
\begin{tikzpicture}
    \small
    \begin{axis}[
            ymajorgrids=true,
            minor y tick num=4,
            extra y ticks={22, 24},
            bar width=.3cm,
            width=1\textwidth,
            height=.3\textwidth,
            legend style={at={(0.5,1.2)},
                anchor=north,legend columns=-1},
             xtick={},
        ]
        \addplot table[x=Segmentation method,y=MuST-C en-de BLEU]{\fixedsegmres};
        \addplot table[x=Segmentation method,y=Europarl-ST en-de BLEU]{\fixedsegmres};
        \addplot table[x=Segmentation method,y=MuST-C en-it BLEU]{\fixedsegmres};
        \addplot table[x=Segmentation method,y=Europarl-ST en-it BLEU]{\fixedsegmres};
        \legend{MuST-C de, Europarl de, MuST-C it, Europarl it}
    \end{axis}
\end{tikzpicture}
\caption{\label{fig:fixed_len_res} BLEU scores (Y axis) with different fixed-length segmentations (in seconds -- X axis).}

\end{figure*}

\noindent \textbf{Fixed-length.}
A simple approach is splitting the audio at a predefined segment length \cite{sinclair_segm}, without considering  the content. In contrast with VAD, this naive method has the benefit of ensuring that the resulting segments are not  too  long or too short, which are typically hard conditions for ST systems. However, the split points are likely to
break
sentences in critical positions  such as between a subject and a verb or even in the middle of a word. Unlike VAD, this method does not filter the
non-speech frames from the input audio, which is entirely passed to the ST system.
Fig. \ref{fig:fixed_len_res} shows that,  with fixed segmentation, translation quality improves with the duration of the segments (slightly for values $>=$16s) up to  20s, after which it decreases.
20 seconds is the maximum segment length in our training data due to memory limits:
we can conclude that longer segments produce better translations, but models can effectively translate only sequences whose length does not exceed the maximum observed in the training set.

\pgfplotstableread[row sep=\\,col sep=&]{
    Segmentation method &
    MuST-C en-de BLEU & MuST-C en-de TER & Europarl-ST en-de BLEU & Europarl-ST en-de TER &
    MuST-C en-it BLEU & MuST-C en-it TER & Europarl-ST en-it BLEU & Europarl-ST en-dt TER \\
    4       & 16.23  & 76.92  & 13.46 & 91.25 & 16.86 & 76.15 & 14.38 & 85.85 \\
    8       & 20.56  & 69.91  & 19.73 & 70.95 & 21.12 & 68.32 & 20.27 & 69.51 \\
    12      & 22.87  & 64.84  & 21.84 & 66.38 & 22.44 & 65.45 & 21.60 & 66.22 \\
    16      & 23.40  & 63.27  & 22.80 & 65.35 & 22.84 & 64.64 & 22.18 & 65.27 \\
    18      & 23.75  & 61.63  & 23.11 & 64.48 & 23.23 & 63.95 & 22.21 & 65.15 \\
    20      & 23.86  & 61.29  & 23.27 & 64.01 & 23.20 & 64.24 & 22.28 & 64.57 \\
    22      & 22.54  & 62.20  & 22.26 & 64.84 & 22.46 & 64.12 & 21.81 & 64.58 \\
}\fixedsegmres

 \noindent \textbf{\mg{SRPOL}-like segmentation.}
The method described in \cite{potapczyk-przybysz-2020-srpols}  takes into account both audio content (silences) and target segments' length (i.e. the desired length of the generated segments) to split the audio.
It recursively divides
the audio segments on the longest silence, until either there are no more silences in a segment, or the segment itself is shorter than a 
threshold. It is important to notice that, in \cite{potapczyk-przybysz-2020-srpols}, the silences are detected with a manual operation,
making the approach hard to 
reproduce
and not scalable. In this paper, we replicate the logic,
but we rely
on WebRTC to automatically identify silences.  For this reason, our results might be slightly different  than the original  ones, but the segmentation is automatic and easy to reproduce. Another major problem of this method is that it requires the full audio to be available for splitting it. So it is not applicable to audio streams and online use cases.
Based on the previous considerations drawn from Fig.\ref{fig:fixed_len_res}, in our experiments  we set the maximum length threshold to 20s, so that the model is fed with sequences that are not longer than the maximum seen at training time. The resulting segments have an average length of 7-8s.

\subsection{Proposed hybrid segmentation}
Similar to \cite{potapczyk-przybysz-2020-srpols}, our method is hybrid
as it considers both the audio content and the target segments' length. However, unlike \cite{potapczyk-przybysz-2020-srpols},  we give more importance to the target segments' length than to the detected pauses  (we motivate this choice in $\S\ref{sec:results}$). Specifically, we split on the longest pause in the interval (minimum and maximum length), if any,  otherwise we split at maximum length. Maximum and minimum segment lengths are controlled by two hyper-parameters (\textit{MAX\_LEN} and \textit{MIN\_LEN}).
Unlike the \mg{SRPOL}-like approach, ours can operate on audio streams, as it does not require the full audio to start the segmentation procedure. Moreover, the latency is controlled by \textit{MAX\_LEN} and \textit{MIN\_LEN}, which can be tuned to trade translation quality for lower latency.

We tested different values for \textit{MIN\_LEN} and we chose 17s for our experiments, because it resulted in the best score on the MuST-C dev set. As in the other methods, and for the same reasons, \textit{MAX\_LEN} is set to  20s.
The resulting segments have an average length slightly higher than 17s.

We also introduce a variant of this method that enforces splitting on pauses  longer than \textit{550ms}. In \cite{karakanta42}, this threshold is shown to often represent a \textit{terminal juncture}: a break between two utterances, usually corresponding to clauses. Splitting on such pauses should hence  enforce separating different clauses. As a result, segments can be shorter than \textit{MIN\_LEN}, but we still ensure  they are not longer than \textit{MAX\_LEN}.
With this variant, the segments are much shorter, as their average length is 8s, similar to the \mg{SRPOL}-like technique.

\begin{table*}[t]
    \centering
    \small
    \begin{tabular}{l|cccccccc}
    \hline
    \textbf{Segm. method} &
    \multicolumn{2}{c}{\textbf{MuST-C en-de}} & \multicolumn{2}{c}{\textbf{Europarl en-de}} &
    \multicolumn{2}{c}{\textbf{MuST-C en-it}} & \multicolumn{2}{c}{\textbf{Europarl en-it}} \\
    & \textbf{BLEU ($\uparrow$)} & \textbf{TER ($\downarrow$)} & \textbf{BLEU ($\uparrow$)} & \textbf{TER ($\downarrow$)} &
    \textbf{BLEU ($\uparrow$)} & \textbf{TER ($\downarrow$)} & \textbf{BLEU ($\uparrow$)} & \textbf{TER ($\downarrow$)} \\
    \hline
    Manual segm. & 27.55 & 58.84 & 26.61 & 60.99 & 27.70 & 58.72 & 28.79 & 59.16 \\
    \hline
    Best VAD & 21.87  & 66.72  & 18.51 & 78.12 & 22.34 & 66.12 & 20.90 & 69.54 \\
    Best Fixed (20s)      & 23.86  & \textbf{61.29}  & 23.27 & 64.01 & 23.20 & 64.24 & 22.28 & 64.57 \\
\mg{SRPOL}-like & 22.26  & 71.10  & 20.49 & 77.61 & 23.12 & 66.27 & 23.26 & 66.19 \\
    \hline
    Pause in 17-20s& \textbf{24.39}  & 61.35  & \textbf{23.78} & \textbf{63.15} & \textbf{23.50} & \textbf{63.76} & 22.86 & 63.44 \\
   \hspace{2mm} + force split & 23.17  & 66.20 & 22.52 & 68.56 & 23.45 & 63.79 & \textbf{24.15} & \textbf{63.31} \\
    \hline
    \end{tabular}
    \caption{\label{tab:custom_segm_res} Comparison between manual and automatic segmentations: VAD, fixed-length and hybrid approaches.}
\end{table*}

\section{Experimental Settings}
\label{sec:expsett}
 
We use a Transformer \cite{transformer}  whose encoder is modified for ST. The encoder starts with two 2D convolutional layers that reduce the length of the  Mel-filter-bank sequence by a factor of 4. The resulting tensors are passed to a linear layer that maps them into the dimension used by the following encoder Transformer layers. A logarithmic distance penalty \cite{di-gangi-etal-2019-adapting} is applied in all the encoder Transformer layers.

The ST models have 11 encoder Transformer layers and 4 decoder Transformer layers. We use 8 attention heads, 512 attention hidden units and 2,048 features in the FFNs' hidden layer. We set dropout to 0.1. The optimizer is Adam \cite{adam} with betas \textit{(0.9, 0.98)}. The learning rate is scheduled with inverse square root decay after 4,000 warm-up updates, during which it increases linearly from $3\cdot10^{-4}$ up to $5\cdot10^{-3}$. The update frequency is set to 8 steps; we train on 8 GPUs and each mini-batch is limited to 12,000 tokens or 8 sentences, so the resulting batch size is slightly lower than 512.
We initialize the convolutional and the first encoder Transformer layers with the encoder of a model trained on ASR data. We pre-train our ST models on the ASR corpora with synthetic targets generated by an MT model fed with the known transcripts \cite{jia2018leveraging} and we fine-tune on the ST corpora. Both these trainings adopt knowledge distillation
\mg{\cite{hinton2015distilling}}
with the MT model as teacher \mg{\cite{liu2019endtoend,gaido-et-al-2020-on-knowledge}}. Finally, we fine-tune on the ST corpora with label-smoothed cross-entropy \cite{szegedy2016rethinking}. In all the three steps, we use SpecAugment \cite{Park_2019} and time stretch \cite{nguyen2019improving} as data augmentation techniques.

The ASR model is similar to ST models, but we use 8 encoder layers and 6 decoder layers. For MT, instead, the Transformer attentions has 16 heads and hidden-layer features are two times those of ST and ASR models.

We experimented with translation from English speech into two target languages: German and Italian. To train the MT model used for knowledge distillation, we employed the WMT 2019 datasets \cite{barrault-etal-2019-findings} and the 2018 release of OpenSubtitles \cite{Lison2016OpenSubtitles2016EL} for English-German and OPUS \cite{opus} for English-Italian. All the data were cleaned with Modern MT \cite{modernmt}. The ASR model, whose encoder was used to initialize that of the ST model, was trained on TED-LIUM 3 \cite{Hernandez_2018}, Librispeech \cite{librispeech}, Mozilla Common Voice\footnote{https://voice.mozilla.org/}, How2 \cite{sanabria18how2}, and the audio-transcript pairs of the ST corpora. The ST corpora were MuST-C \cite{CATTONI2021101155} and Europarl-ST \cite{europarlst} for both target languages. We filtered out samples with input audio longer than 20s to avoid out-of-memory errors. \mg{The text is encoded using BPE \cite{sennrich2015neural} with 8,000 merge rules \cite{di-gangi-etal-2020-target}.}

\pgfplotstableread[row sep=\\,col sep=&]{
    Segmentation method &
    MuST-C en-de & Europarl-ST en-de & MuST-C en-it & Europarl-ST en-it \\
     4      &  2.18672912   & 2.33661851    &  1.662943637    &  2.325718253  \\
     8      &  0.0832604441 & 0.1472073931  &	 0.7943382636 &  0.2294636603 \\
    12      & -0.1656292101 & -0.1743604969 &	 0.3749600679 & -0.2345313207 \\
    16      & -0.2147829662 & -0.3351444419 & -0.05934261878  & -0.3631823246 \\
    18      & -0.2834421812 & -0.488158159  &	-0.4249926471 & -0.4951441999 \\
    20      & -0.1835742322 & -0.6142749337 &	-0.6627895363 & -0.5949433243 \\
    22      & -1.422560975  & -0.8718878715 & -1.685117166    & -0.8673807443 \\
}\fixedsegmlen

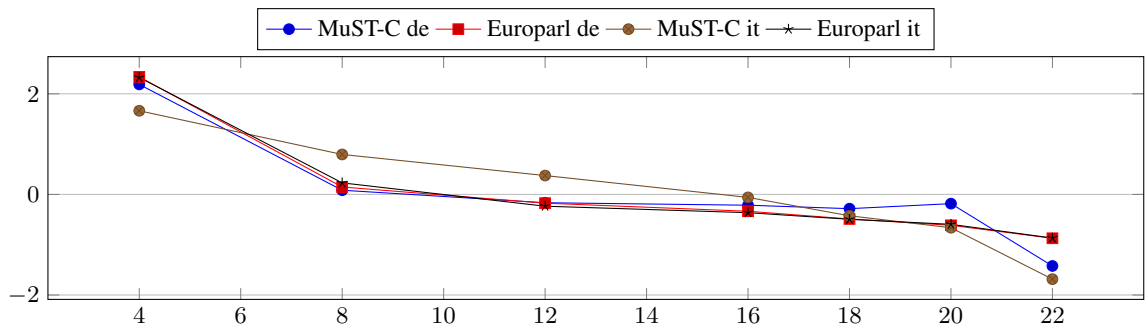
\begin{figure*}[!ht]
\centering
\begin{tikzpicture}
    \small
    \begin{axis}[
            ymajorgrids=true,
            width=1\textwidth,
            height=.3\textwidth,
            legend style={at={(0.5,1.2)},
                anchor=north,legend columns=-1},
            xtick=data,
            xtick={},
        ]
        \addplot table[x=Segmentation method,y=MuST-C en-de]{\fixedsegmlen};
        \addplot table[x=Segmentation method,y=Europarl-ST en-de]{\fixedsegmlen};
        \addplot table[x=Segmentation method,y=MuST-C en-it]{\fixedsegmlen};
        \addplot table[x=Segmentation method,y=Europarl-ST en-it]{\fixedsegmlen};
        \legend{MuST-C de, Europarl de, MuST-C it, Europarl it}
    \end{axis}
\end{tikzpicture}
\caption{\label{fig:output_len} Z-score normalized output lengths (number of words) according to the input segments length.}
\end{figure*}

{
\begin{table*}[h!]
  \centering
  \setlength{\tabcolsep}{1.5pt}
  \footnotesize
  \begin{tabular}{p{1.5cm}|p{14cm}}
      \hline
      \multicolumn{2}{c}{(a) Hallucinations with non-speech audio} \\
      \hline
      \textbf{Audio} & \textit{Music and applause}. \\
      \textbf{4s segments} & \textbf{[Chinesisch] [Hawaiianischer Gesang] // Chris Anderson: Du bist ein Idiot. // Nicole: Nein.}
      
      \textit{[Chinese] [Hawaiian song] // Chris Anderson: You are an idiot. // Nicole: No.}
      \\
      \hline
      \multicolumn{2}{c}{(b) Hallucinations with sub-sentential utterances} \\
      \hline
      \textbf{Audio} & Now, chimpanzees are well-known for their aggression. // (Laughter) // But unfortunately, we have made too much of an emphasis of this aspect (...) \\
      \textbf{Reference} & Schimpansen sind bekannt für ihre Aggressivität. // (Lachen) // Aber unglücklicherweise haben wir diesen Aspekt überbetont (...) \\
      \textbf{4s segments} & \textbf{Publikum: Nein.} Schimpansen sind bekannt. // \textbf{Ich bin für} ihre Aggression gegangen. // Aber leider haben \textbf{wir zu viel Coca-Cola gemacht}. // \textbf{Das ist eine} wichtige Betonung dieses Aspekts (...)
      
      \textit{\textbf{Audience: No.} Chimpanzees are known. // \textbf{I went for} their aggression. // But unfortunately \textbf{we made too much Coca-Cola}. // \textbf{This is an} important emphasis of this aspect (...)}
      \\
      \textbf{20s segments} & Schimpansen sind bekannt für ihre Entwicklung. //
Aber leider haben wir zu viel Schwerpunkt auf diesem Aspekt (...)
      
      \textit{Chimpanzees are known for their development. // But unfortunately, we have expressed too much emphasis on this aspect (...)}
      \\
      \hline
      \multicolumn{2}{c}{(c) Hallucinations and bad translation with sub-sentential utterances} \\
      \hline
      \textbf{Audio} & (...) where the volunteers supplement a highly skilled career staff, you have to get to the fire scene pretty early to get in on any action. \\
      \textbf{Reference} & (...) in der Freiwillige eine hochqualifizierte Berufsfeuerwehr unterstützten, muss man ziemlich früh an der Brandstelle sein, um mitmischen zu können. \\
      \textbf{4s segments} &  (...) \textbf{wo die Bombenangriffe auf dem Markt waren. // Man muss bis zu 1.000 Angestellte in die USA, nach Nordeuropa kommen}.
      
      \textit{(...) \textbf{where the bombings were on the market. // You have to come up to 1,000 employees in the USA, to Northern Europe}.}
      \\
      \textbf{20s segments} & (...) in der die Freiwilligen ein hochqualifiziertes Karriere-Team ergänzen, muss man ziemlich früh an die Feuerszene kommen, um in irgendeiner Aktion zu gelangen.
      
      \textit{(...) where the volunteers complement a highly qualified career team, you have to get to the fire scene pretty early in order to get into any action.}
      \\
      \hline
      \multicolumn{2}{c}{(d) Final portions of long segment ignored} \\
      \hline
      \textbf{Audio} & But still it was a real footrace against the other volunteers to get to the captain in charge to find out what our assignments would be. // When I found the captain, (...)\\
      \textbf{Reference} & Aber es war immer noch ein Wettrennen gegen die anderen Freiwilligen \textbf{um den verantwortlichen Hauptmann zu erreichen und herauszufinden was unsere Aufgaben sein würden}. // Als ich den Hauptmann fand (...)\\
      \textbf{22s segments} & (...) Es war immer noch ein echtes Fussrennen gegen die anderen Freiwilligen. // Als ich den Kapitän fand, (...)
      
      \textit{(...) It was still a real footrace against the other volunteers. // When I found the captain, (...)}
      \\
      \hline
  \end{tabular} 
  \caption{\label{tab:analysis} Translations affected by errors caused by too short -- \textit{(a)}, \textit{(b)}, \textit{(c)} -- or too long -- \textit{(d)} --  segments. The symbol ``//'' refers to a break between two segments. The breaks might be located in different positions in the different segmentations.
  Over-generated -- in examples \textit{(a)}, \textit{(b)}, \textit{(c)} -- and missing -- in \textit{(d)} -- content is marked in \textbf{bold} respectively in system's outputs and in the reference.}
\end{table*}
}

\section{Results}
\label{sec:results}
We compute ST results in terms of BLEU \cite{papineni-2002-bleu} and TER \cite{snover-ter-2006} on the MuST-C and Europarl-ST test sets for en-de and en-it. In MuST-C the two test sets are identical with regard to the audio, while in Europarl-ST they contain different recordings.

As shown in Table \ref{tab:custom_segm_res},  fixed-length segmentation always outperforms  the best VAD, both in terms of BLEU and TER. 
This may be surprising but it confirms previous findings in ASR  \cite{sinclair_segm}: also in ST, VAD is more costly and less effective than a naive fixed-length segmentation. Besides, it suggests that the resulting segments' length is more important than the precision of the split times. This observation motivates the definition of our proposed techniques.
Compared to fixed-length segmentation,
the \mg{SRPOL}-like method provides better results for en-it, but worse for en-de, indicating that the syntactic properties of the source and target languages are an important factor for audio segmentation (see $\S$\ref{sec:analysis}).

Our proposed method (\textit{Pause in 17-20s} in Table \ref{tab:custom_segm_res})
outperforms the others on all test sets but Europarl en-it, in which
\mg{SRPOL}-like
has a higher BLEU (but worse TER).
The version with forced splits on 550ms pauses
is inferior  to the version without forced splits  on the German test sets, but it is on par for MuST-C en-it and superior on Europarl en-it, on which it is the best segmentation overall by a large margin. Moreover, its scores are always better than the ones obtained by the \mg{SRPOL}-like approach, although the length of the produced segments is similar.
These results suggest  that, although the best version depends on the syntax and the word order of the source and target languages, our method can always outperform the others in terms of both BLEU and TER. Noticeably, it does not introduce latency, since it does not require the full audio to be available for splitting it, as the \mg{SRPOL}-like technique does. In particular, averaged on the two domains, our best results (respectively with and without \textit{forced splits}) 
reduce the gap with the manual segmentation by 54.71\% (en-de) and 30.95\% (en-it) compared to VAD-based segmentation.

\section{Analysis}
\label{sec:analysis}

A first interesting consideration regards the length of the produced translations.
In particular, we analyze the case of fixed-length segmentation
(see Fig. \ref{fig:output_len}): in presence of short input segments the output is longer, while it gets shorter in case of segments longer than  20s.
To understand this behavior, we performed a manual inspection of the German translations produced by fixed-length segmentation with 4s, 20s and 22s.

The analysis revealed two main types of errors: overly long  (\textit{hallucinations} \cite{Lee2018HallucinationsIN}) and overly short outputs.
The first type of error occurs when the system is fed with  small, sub-sentential segments. In this case, trying to generate well-formed sentences, the system ``completes'' the translation with text that has no correspondence with the input utterance.
The second type of error occurs when the system is fed with segments that exceed the maximum length observed in the training data. In this case, part of the input  (even complete clauses, typically towards the end of the utterance) is not realized in the final translation.

Table \ref{tab:analysis} provides  examples of all these phenomena.
The first three examples show cases of hallucinations in short (4s) segments, while the last one shows an incomplete translation of a long (22s) segment. In particular:

 \textit{(a)} shows the generation of text not related to the source when the audio contains only noise or silence (e.g. at the beginning of a TED talk recording).

 \textit{(b)} presents the addition of non-existing content in the translation of a sub-sentential segment.

 \textit{(c)} is related to a sub-sentential utterance as well, but in this case the output of the system is affected by both hallucinations and poor translation quality due to the lack of enough context.

 \textit{(d)} reports a segment whose last portion
 ignored.

The length of the generated outputs also helps understanding the different results obtained by the variants of our method on the two target languages. Indeed, the introduction of forced splits (\textit{+ force split}) produces audio segments that are much shorter (\texttildelow 8s vs \texttildelow 17s) and hence, according to the previous consideration, the resulting translation is overall longer. For German, the difference in terms of output length is high ($>$ 8.5\%), while for Italian it is much lower (4.33\% on MuST-C and 2.49\% on Europarl-ST). So, the German results are penalized by the additional  hallucinations, while, for Italian translations, the beneficial separation of clauses delimited by terminal juncture dominates.

This different behavior relates to the different syntax of the source and target languages. Indeed, translating from English (an SVO language) into German (an SOV language) requires long-range re-orderings 
\cite{gojun-fraser-2012}, which can also span over  sub-clauses. The Italian phrase structure, instead, is more similar to English. This is confirmed by the shifts counted in TER computation, which are 20\% more in German than in Italian. Moreover, in Italian their number does not change between our  method with and without forced splits, while in German the version with forced splits has 5-10\% more shifts.

\section{Conclusions}

We studied  different segmentation techniques for direct ST. Despite its  wide adoption, VAD-based segmentation resulted to be underperforming.  We showed that  audio segments' length is a crucial factor to obtain good translations and that the best segmentation approach depends on the structural similarity between the source and target languages. In particular, we demonstrated that the resulting segments should be neither longer than the maximum length of the training samples nor too short (especially when the target language has a different structure).  Inspired by these findings, we proposed two variants of a hybrid method  that significantly improve on different test sets and languages over the VAD baseline and the other techniques presented in literature. Our approach was designed to be also applicable to audio streams and to allow controlling latency, hence being suitable  even for online use cases.

\bibliographystyle{acl_natbib}
\bibliography{anthology,acl2021}


\end{document}